\documentclass[a4paper,11pt]{article}

\usepackage{bbm}
\usepackage{booktabs}
\usepackage{caption}
\usepackage{color}
\usepackage{dsfont}
\usepackage{gensymb}
\usepackage{graphicx}
\usepackage{hepnames}
\usepackage{physics}
\usepackage{pos}
\usepackage[separate-uncertainty=true]{siunitx}
\usepackage{upgreek}
\usepackage{url}
\usepackage{xpatch}
\usepackage{xspace}

\usepackage[capitalize]{cleveref}

\definecolor{LightCyan}{rgb}{0.88,1,1}
\definecolor{Pink}{rgb}{1,0.8, 0.7}

\makeatletter
\xpatchcmd\@HepConStyle
 {\edef\@upcode{\updefault}}
 {\ifdefined\shapedefault\edef\@upcode{\shapedefault}\else\edef\@upcode{\updefault}\fi}
 {}{}
\makeatother

\renewcommand{\Prho}{\ensuremath{\uprho}\xspace}
\renewcommand{\pi}{\uppi}
\renewcommand{\tau}{\uptau}
\renewcommand{\nu}{\upnu}
\newcommand{\Paia}{\HepParticleResonanceFull{a}{1}{}{1420}{}{}\xspace}
\newcommand{\Paib}{\HepParticleResonanceFull{a}{1}{}{1640}{}{}\xspace}
\renewcommand{\Pfz}{\HepParticleResonanceFull{f}{0}{}{980}{}{}\xspace}

\newcommand{\tautothreepi}{\HepProcess{\Ptauon\to\Ppiminus\Ppiplus\Ppiminus\Pnut}\xspace}
\newcommand{\tautothreepigen}{\HepProcess{\Ptauon\to\Ppi\Ppi\Ppi\Pnut}\xspace}

\newcommand{\PartialWaveBox}[2]{\ensuremath{[#1 \Ppi]_{\text{#2}}}\xspace}
\newcommand{\PartialWave}[4]{\ensuremath{#1^{#2}\PartialWaveBox{#3}{#4}}\xspace}
\newcommand{\BDT}{BDT\xspace}
\newcommand{\CM}{CM\xspace}
\newcommand{\PWA}{PWA\xspace}
\newcommand{\QMIPWA}{QMIPWA\xspace}

\DeclareCaptionLabelFormat{tableandfigure}{#1~#2 \& \figurename~\thefigure}

\title{Partial wave analysis of \tautothreepi at Belle}
\ShortTitle{PWA of $\tau\to3\pi\nu_\tau$ at Belle}

\author*[a]{Andrei Rabusov}
\author[a]{Daniel Greenwald}
\author[a]{Stephan Paul}

\affiliation[a]{Technical University of Munich,\\
  James-Franck str.\ 1, Garching 85648, Germany}

\emailAdd{a.rabusov@tum.de}

\abstract{

  We present simulation studies in preparation for analyzing
  \tautothreepi in data from the Belle experiment at the KEK
  $\APelectron\Pelectron$ collider. Analyzing this decay can shed
  light on the \Pai and \Paia resonances and yield results that
  improve measurement of the \Ptau electric and magnetic dipole
  moments. We show that we can achieve a higher signal efficiency than
  previous analyses of the same decay. We also demonstrate that neural
  networks can model our complicated six-dimensional background
  distributions and that quasi-model-independent partial-wave analysis
  can extract resonance masses, widths, and production amplitudes and
  phases.

}

\FullConference{%
  41st International Conference on High Energy physics --- ICHEP2022\\
  6-13 July, 2022\\
  Bologna, Italy
}

\begin{document}
\maketitle

In the decay \tautothreepi, hadrons are produced from unflavored
axial-vector resonances~\cite{pluto}. This is an opportune setting in
which to study such composite particles without strong interaction
with other particles that may alter their resonance shapes. The
dominantly produced resonance is the \Pai, whose shape is much debated
and whose mass and width are not well
determined~\cite{compass-a1,cleo-ii-tau3pi,PDG2020}. The COMPASS
experiment observed an unexpected narrow axial-vector resonance,
\Paia, in partial-wave analysis~(\PWA) of three-pion final states
produced in pion-proton scattering~\cite{compass-a1-1420}. Whether
this is a true particle resonance or an effect of $\PKstar\PK$
scattering is debated~\cite{balalaika}.

A better model for \tautothreepi, driven by experimental measurement,
will improve the simulation of this decay in existing MC generators,
which is necessary for general \Ptau studies at currently running
experiments such as Belle~II~\cite{b2pb}. In particular, it will improve
measurement of the tauon electric and magnetic dipole
moments~\cite{krinner-edm}.

The Belle experiment, which ran for a decade at the
\num{10.58}-\si{GeV} $\APelectron\Pelectron$ collider KEKB in Tsukuba,
Japan, can study the \Pai and \Paia and the general structure of
\tautothreepi using partial-wave analysis and data containing
\num{50e6} \tautothreepi decays~\cite{belle}. This data size is
comparable to that of the COMPASS experiment, five and fifty times
larger than what the Belle and Babar experiments used to publish
$\Ppi\Ppi\Ppi$ mass spectra, and one-thousand times larger than what
the CLEO~II experiment used to publish the only amplitude analysis of
\tautothreepigen~\cite{compass-a1-1420,lee,nugent,cleo-ii-tau3pi}.

We present preliminary studies of the applicability of \PWA to
\tautothreepi using simulated data.  Since Belle cannot detect
neutrinos and \Ptau decays are measured in events with at least two
neutrinos, we do not know the full coordinate of each decay in its
eight-dimensional phase space. We analyze in a six-dimensional
subspace spanned by the three-pion mass, $m_{3\Ppi}$, the two
$\Ppiplus\Ppiminus$ squared masses, $s_1$ and $s_2$, and the three
Euler angles, $\alpha$, $\beta$, and $\gamma$, defined in
\cite{kuehn1992}. We average decay rates over the unknown neutrino
direction and calculate them from hadronic currents written in the
relativistic tensor formalism of~\cite{fkrinner:tensor}.

We study data simulated as if it is produced by the Belle experiment,
with all known interactions originating from $\APelectron\Pelectron$
collision, including $\APelectron\Pelectron\to\APtauon\Ptauon$.  To
isolate $\APtauon\Ptauon$ events containing \tautothreepi, we select
events that each have four charged particles, having total charge
zero, coming from the $\APelectron\Pelectron$ interaction region, each
with transverse momentum in the lab frame above \SI{100}{MeV}. We
select events with a $3{\times}1$ topology relative to the thrust axis
in the $\APelectron\Pelectron$ center-of-momentum~(\CM) frame.

We use a boosted-decision-tree algorithm (\BDT) from the \texttt{ROOT}
\texttt{TMVA} library to further select signal decays and veto
background events; it looks at six event-wide kinematic variables.
After selecting events by their \BDT score, we further select for
pion-identification quality and veto events in which any pair of
oppositely charged pions are consistent with coming from a \PKshort or
in which the total energy of photons in the signal hemisphere is
consistent with the presence of one or more \Ppizero. Photons counted
for the veto must have energy above \SI{40}{MeV} in the lab frame. Our
signal efficiency is \num{31}\%, with a signal purity of \num{87}\%.

\begin{figure}[t]
    \begin{centering}
        \includegraphics[width=0.85\textwidth]{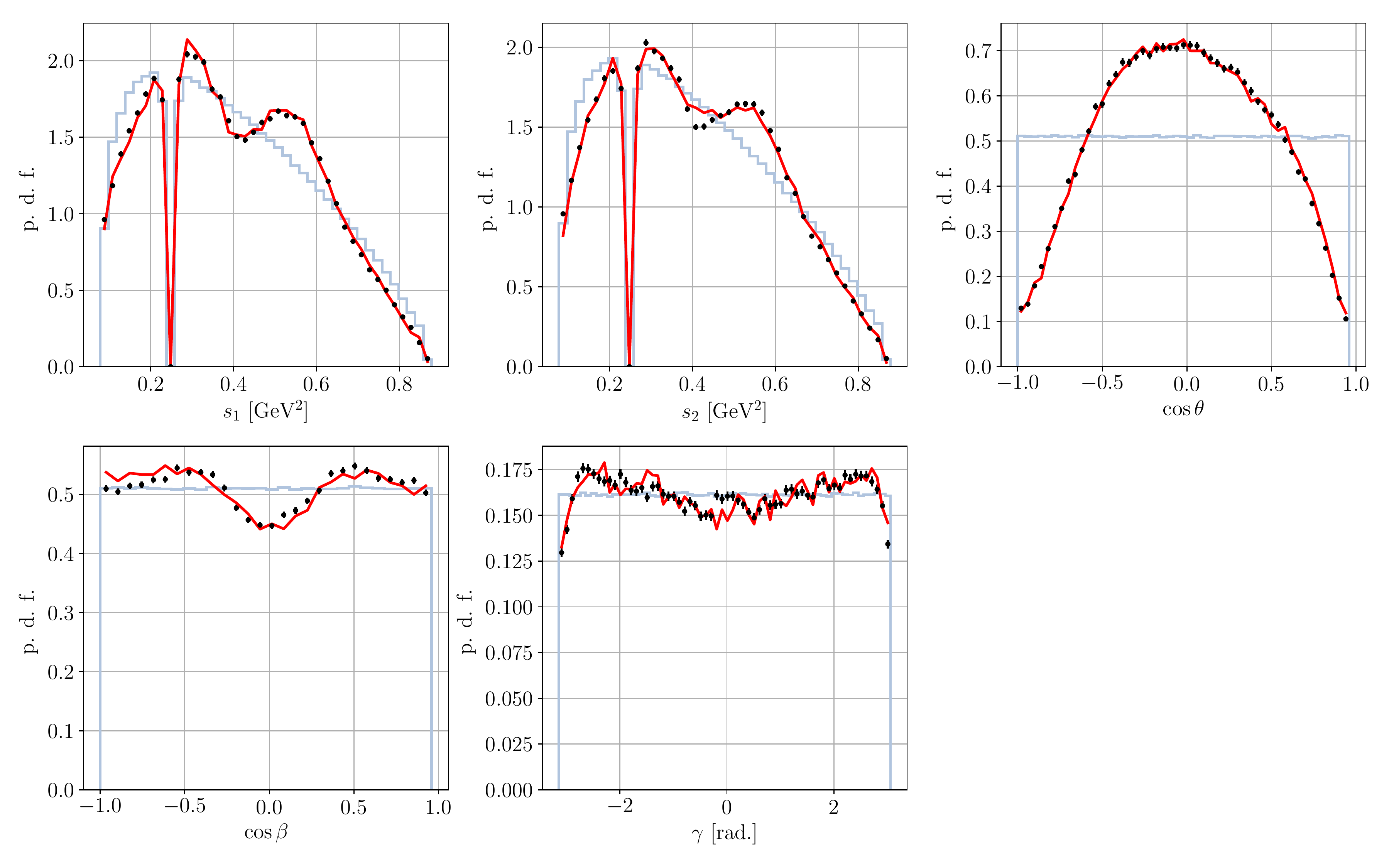}
        \caption{Distribution in simulation~(black), from
          neural-network~(red), and structureless~(blue)}
        \label{fig:nn-proj}
    \end{centering}
\end{figure}

The other \num{13}\% of events are from
\HepProcess{\APelectron\Pelectron\to\APtauon\Ptauon} in which the
three-prong tau decay is
\HepProcess{\Ptauon\to\Ppiminus\Ppiplus\Ppiminus\Ppizero\Pnut} (with
possible further \Ppizero) or
\HepProcess{\Ptauon\to\PKminus\Ppiplus\Ppiminus\Pnut} or from
\HepProcess{\APelectron\Pelectron\to\Pquark\APquark}. The dynamic
structure of these backgrounds in the 6D analysis space is too
complicated to model parametrically. Instead we let a neural network
learn the background shape, a method pioneered in amplitude analysis
by LHCb in~\cite{polu-nn} to use a single neural network to
parametrize the background in the entire phase space. We find it
necessary to train multiple neural networks, each for a subregion of
$m(3\pi)$. \cref{fig:nn-proj} shows the resulting background shape for
$m_{3\Ppi} \in \SIrange{1.06}{1.08}{GeV}$.  The neural network
prediction agrees with the simulated background.

We analyze the data in subregions of $m_{3\pi}$ with background
modeled by the neural network and signal modeled with isobars and
quasi-model-independent partial-wave analysis (\QMIPWA) as described
in~\cite{zm-paper}. To cross check the method, we analyze data
simulated with only four partial waves: \PartialWave{1}{+}{\Pfz}{P},
\PartialWave{1}{+}{\Prho(770)}{S}, \PartialWave{1}{+}{\Prho(770)}{D},
and \PartialWave{1}{+}{\Pfii}{P}. We use a \QMIPWA isobar for the
\PartialWave{1}{+}{1^{--}}{S} wave only, to avoid zero modes and
simplify the test. We fit the \QMIPWA complex amplitudes and a complex
multiplier for each remaning wave. We then fit a Breit-Wigner function
to the \QMIPWA results~(Fig.~2). The second fit determines the \Prho's mass and
width to be \SI{769.8 +- 0.6}{MeV} and \SI{155.2 +- 1.3}{MeV},
agreeing with the simulated values of \SI{769.0}{MeV} and
\SI{150.9}{MeV}.  The fit results (\cref{tab:freed-io-test}) all agree
with their simulated values.

\begin{table}[t]
  \begin{minipage}{0.64\textwidth}
    \begin{tabular}{@{}l@{\hspace{0.5em}}*{2}{rr@{$\,\pm\,$}r}@{}}
      \toprule

      \sisetup{round-mode=places,round-precision=2}
      
      Wave
      & \multicolumn{3}{c}{Amplitude}
      & \multicolumn{3}{c}{Phase [deg]}
      \\
      
      \cmidrule{2-4}
      \cmidrule(l){5-7}
      
      & {sim.}
      & \multicolumn{2}{c}{res.}
      & {sim.}
      & \multicolumn{2}{c}{res.}
      \\

      \midrule

      \PartialWaveBox{\Pfz}{P}
      & \num{0.10}
      & \num{0.099} & \num{0.001}
      & \num{-60}
      & \num{-55.485} & \num{1.947}
      \\

      \PartialWaveBox{\Prho(770)}{S}
      & \num{0.70}
      & \num{0.712} & \num{0.005}
      & \num{0}
      & \multicolumn{2}{c}{reference phase}
      \\

      \PartialWaveBox{\Prho(770)}{D}
      & \num{0.92}
      & \num{0.959} & \num{0.025}
      & \num{120}
      & \num{120.546} & \num{0.459}
      \\

      \PartialWaveBox{\Pfii}{P}
      & \num{0.53}
      & \num{0.514} & \num{0.020}
      & \num{15}
      & \num{18.255} & \num{2.525}
      \\

      \bottomrule
    \end{tabular}
  \end{minipage}%
  \begin{minipage}{0.36\textwidth}
    \includegraphics[width=\textwidth]{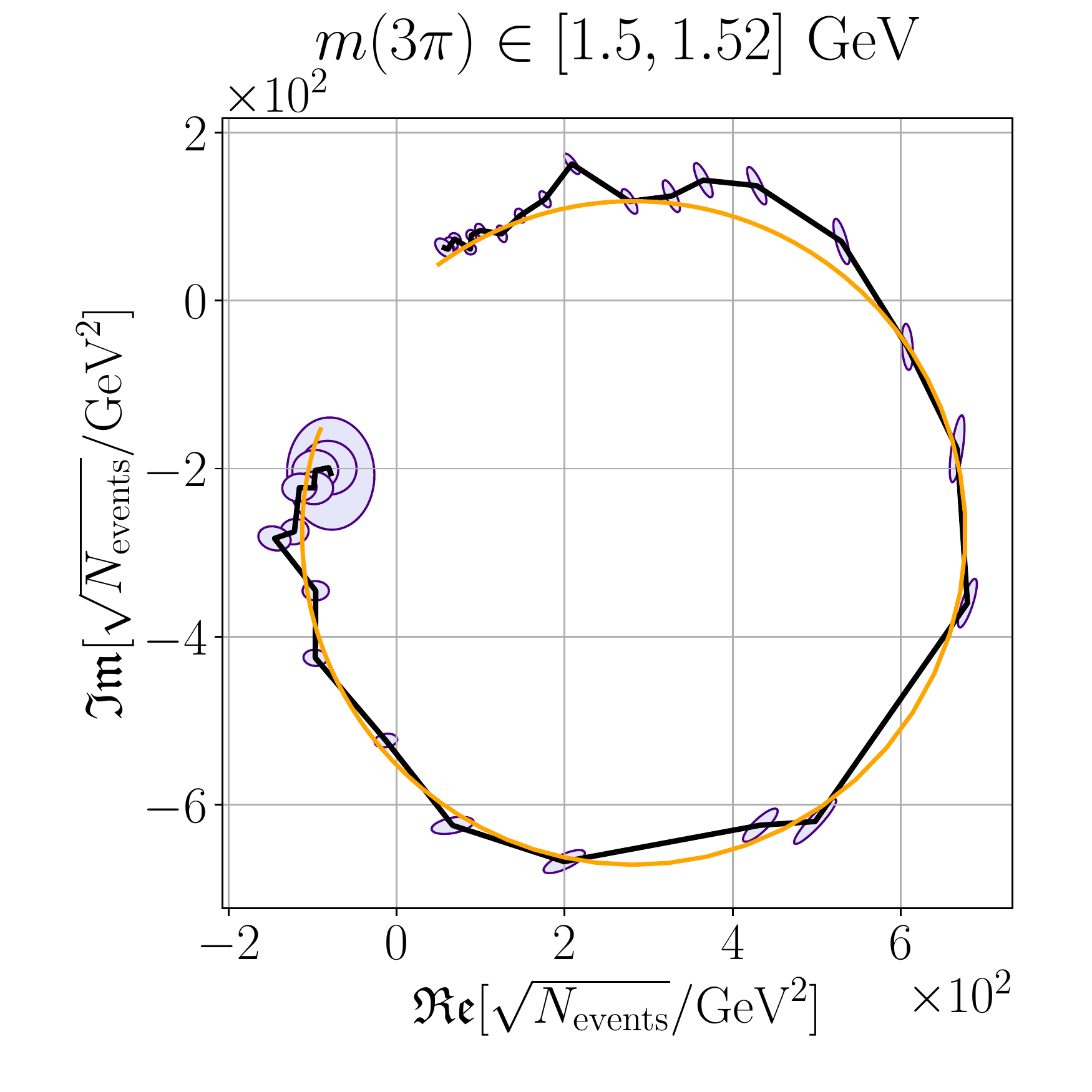}
  \end{minipage}
  
  \caption{
    Comparison of simulated values and fit results.
    {\bf Figure 2:} \QMIPWA~(violet) and Breit-Wigner~(orange) fit
    results for the \PartialWave{1}{+}{1^{--}}{S} wave in simulated
    data; elipses show 68\%-confidence intervals.
    \label{tab:freed-io-test}
  }
\end{table}

In conclusion, we have developed selection criteria with higher
efficiency than previously achieved by the BaBar and Belle
experiments~\cite{nugent,lee}, though with a higher background
contamination. However, we can still analyze this data well using a
neural-network to parameterize background. We have also demonstrated
that a fit algorithm using quasi-model-independent partial-wave
analysis reproduces simulation inputs. This technique will be useful
to study the \Pai, \Paia, \Paib and general structure of \tautothreepi
independent of a model.

\bibliographystyle{JHEP}
\bibliography{db}

\end{document}